\documentclass[prl,floatfix,showpacs,twocolumn]{revtex4}

\usepackage{amsmath}
\usepackage{amssymb}
\usepackage{graphicx}
\usepackage{times}

\newcommand{\eg}{\emph{e.}$\,$\emph{g.}}
\newcommand{\ie}{\emph{i.}$\,$\emph{e.}}
\newcommand{\etal}{\emph{et}$\,$\emph{al.}}

\newcommand{\romc}{{\operatorname{c}}}
\newcommand{\rome}{{\operatorname{e}}}
\newcommand{\romf}{{\operatorname{f}}}
\newcommand{\romi}{{\operatorname{i}}}
\newcommand{\romr}{{\operatorname{r}}}
\newcommand{\romB}{{\operatorname{B}}}

\newcommand{\VECn}{{\boldsymbol{n}}}
\newcommand{\VECq}{{\boldsymbol{q}}}
\newcommand{\VECr}{{\boldsymbol{r}}}
\newcommand{\VECa}{{\boldsymbol{a}}}

\begin{document}

\title{Efficient tunable generic model for fluid bilayer membranes}

\author{Ira R. Cooke}
\author{Kurt Kremer}
\author{Markus Deserno}

\affiliation{Max-Planck-Institut f\"ur Polymerforschung, %
             Ackermannweg 10, %
             55128 Mainz, %
             Germany}

\date{\today}
\begin{abstract}
  We present a model for the efficient simulation of generic bilayer
  membranes.  Individual lipids are represented by one head- and two
  tail-beads.  By means of simple pair potentials these robustly
  self-assemble to a fluid bilayer state over a wide range of
  parameters, \emph{without} the need for an explicit solvent.  The
  model shows the expected elastic behavior on large length scales,
  and its physical properties (\eg\ fluidity or bending stiffness) can
  be widely tuned via a single parameter.  In particular, bending
  rigidities in the experimentally relevant range are obtained, at
  least within $3-30 k_{\text{B}}T$.  The model is naturally suited to
  study many physical topics, including self-assembly, fusion, bilayer
  melting, lipid mixtures, rafts, and protein-bilayer interactions.
\end{abstract}


%
%
%

\maketitle


Lipid molecules in aqueous solution spontaneously assemble into
bilayer membranes.  In biological systems, such membranes are involved
in tasks over an extraordinary range of length scales, from transport
of water and ions at the scale of nm, up to phagocytosis, amoebal
motion and cell budding at the scale of $\mu $m \cite{Lodish}.
Computer simulations designed to understand some aspects of this
structural and functional range must accordingly be tailored to the
specific length and timescales involved.  Techniques that probe both
the smallest \cite{atomistic_mem} and largest \cite{GoKr:04} of these
length and time scales are now comparatively well established;
however, accessing intermediate regimes has proven far more difficult,
and it is only recently that significant progress has been made in
this regard.  The need for a comprehensive suite of techniques to
study lipid bilayers at the mesoscale is highlighted by the sheer
number of relevant problems in this regime, which include viral
budding, raft formation, fusion, phase separation of multicomponent
systems and protein sorting during vesiculation.

Most existing approaches to mesoscale simulation employ coarse grained
lipids and require explicit solvent particles to stabilize the
bilayer.  This strategy is convenient and natural, yet it comes at a
high price: Already for small flat systems the solvent accounts for
most of the computational effort, but the problem gets significantly
worse when three dimensional objects such as vesicles are to be
simulated.  Membrane and solvent play the role of surface and bulk,
respectively, hence the solvent degrees of freedom vastly outnumber
the lipids even for rather modest sized vesicles.  An obvious solution
to this problem is to replace the solvent by effective lipid
interactions.  Given the great success of this approach in polymer
physics it is perhaps surprising that solvent-free bilayer simulations
have so far failed to find widespread acceptance.  The problem appears
to be that naive choices for interparticle potentials (\eg\
Lennard-Jones) do not lead to a fluid bilayer phase but only to
ordered ``solid'' bilayers at low temperature and low density phases
at high temperature.  Many attempts to obtain a broadly stable fluid
phase have been made with varying degrees of success.  So far,
however, none of these has resulted in a model that is sufficiently
robust, simple, or versatile for general use.  For example, the
original solvent-free model of Drouffe \etal\ \cite{DrMa91} and later
modifications by Noguchi \cite{Noguchi} rely on density dependent
interaction potentials to stabilize the fluid phase.  But the
multibody nature of interactions poses serious problems for
interpretation and measurement of thermodynamic quantities.  Other
models do not exhibit the crucial property of unassisted self-assembly
\cite{Oded,BrBr04} and require the use of angular dependent potentials
\cite{BrBr04} or a large set of highly tuned interaction parameters
\cite{Oded}.  Thus, there remains a clear need for an efficient,
robust and tuneable solvent-free bilayer model.

In this letter we present a very simple method for simulating
solvent-free fluid bilayer membranes that is based on pair
potentials, is highly robust and tuneable, and reproduces
macroscopic properties of real bilayers.  The key ingredient is an
attractive potential between lipid tails for which the
\emph{range} of attraction, $w_\romc$, can be varied.  First we
map out the properties of a lipid system as a function of
$w_\romc$ and temperature and find that a sufficiently large
$w_\romc$ is the key to obtaining a fluid phase. Moreover,
$w_\romc$ can be used to tune bilayer properties and to construct
multi-component systems with interfacial tension.  This is
demonstrated in the final part of our Letter where we examine the
kinetics of domain formation in a two-component vesicle.

Let us now describe a model that demonstrates the principle mentioned
above.  Each lipid molecule is represented by one ``head'' bead
followed by two ``tail'' beads.  Their size is fixed via a
Weeks-Chandler-Andersen potential
\begin{equation}
  V_{\text{rep}}(r;b) = \left\{
  \begin{array}{c@{\;\;,\;\;}c}
    4\epsilon\big[(\frac{b}{r})^{12}-(\frac{b}{r})^6+\frac{1}{4}\big] & r \le r_\romc \\
    0 & r>r_\romc
  \end{array}
  \right. \ ,
\end{equation}
with $r_\romc=2^{1/6}b$.  We use $\epsilon$ as our unit of energy.  In
order to ensure an effective cylindrical lipid shape we choose
$b_{\text{head,head}} = b_{\text{head,tail}}=0.95\,\sigma$ and
$b_{\text{tail,tail}}=\sigma$, where $\sigma$ is the unit of
length. The three beads are linked by two FENE bonds
\begin{equation}
  V_{\text{bond}}(r) = -\textstyle\frac{1}{2}k_{\text{bond}}\,r_\infty^2\log\big[1-(r/r_\infty)^2\big] \ ,
\end{equation}
with stiffness $k_{\text{bond}}=30\,\epsilon/\sigma^2$ and divergence
length $r_\infty=1.5\,\sigma$.  Lipids are straightened by a harmonic
spring with rest length $4\sigma$ between head-bead and second
tail-bead
\begin{equation}
  V_{\text{bend}}(r) = \textstyle\frac{1}{2}k_{\text{bend}}(r-4\sigma)^2 \ ,
\end{equation}
which corresponds in lowest order to a harmonic bending potential
$\frac{1}{2} k_{\text{bend}}\sigma^2\,\vartheta^2$ for the angle
$\pi-\vartheta$ between the three beads.  We fixed the bending
stiffness at $k_{\text{bend}}\sigma^2=10\,\epsilon$.  Finally, all
\emph{tail} beads attract each other according to
\begin{equation}
  V_{\text{attr}}(r) = \left\{
  \begin{array}{c@{\;\;,\;\;}c}
    -\epsilon & r < r_\romc \\
    -\epsilon\,\cos^2\frac{\pi(r-r_\romc)}{2w_\romc} & r_\romc \le r \le r_\romc + w_\romc \\
    0 & r > r_\romc+w_\romc
  \end{array}
  \right. \ ,
\label{eqn:pot}
\end{equation}
This describes an attractive potential with a depth of $\epsilon$
which for $r>r_\romc$ smoothly tapers to zero.  Its decay range
$w_\romc$ is the key tuning parameter in our model.

We performed Molecular Dynamics (MD) simulations using the
ESPResSo package \cite{espresso}.  A fixed number of lipids was
simulated in a cubic box of side length $L$ subject to periodic
boundary conditions.  The canonical state was reached by means of
a Langevin thermostat \cite{GrKr86} (with a time step $\delta
t=0.01\,\tau$ and a friction constant $\Gamma=\tau^{-1}$ in
Lennard-Jones units).  If needed, constant tension conditions were
also implemented via a modified Andersen barostat \cite{KoDu99}
(with a box friction $\Gamma_{\text{box}} = 2 \times
10^{-4}\,\tau^{-1}$ and box mass within the range $Q = 10^{-5}
\ldots \times 10^{-4}$).

\begin{figure}
\includegraphics[scale=0.32]{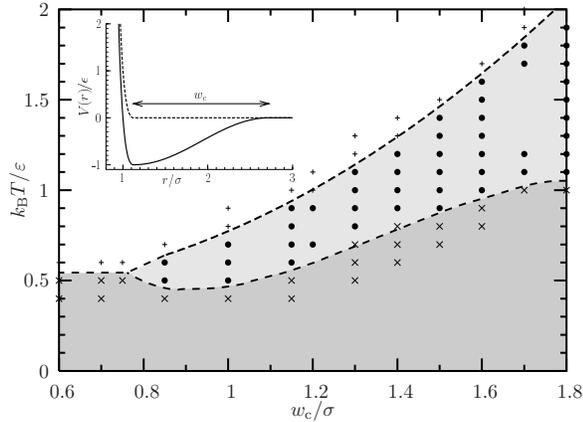}
\caption{Phase diagram in the plane of potential
  width $w_\romc$ and temperature at zero lateral tension.  Each
  symbol corresponds to one simulation and identifies different
  bilayer phases: $\times$: gel; $\bullet$: fluid, {\footnotesize
  $+$}: unstable.  The dashed lines are merely guides to the eye.  The inset
  shows the pair-potential between tail lipids (solid line) and the
  purely repulsive head-head and head-tail interaction (dashed
  line).}\label{fig:pd}
\end{figure}

One of our principal objectives is to map out the conditions under
which the fluid bilayer is stable.  We identified the fluid phase in
two different ways.  First, a box-spanning bilayer was pre-assembled
from 1000 lipids, and its equilibration under zero lateral tension was
attempted (requiring -- if successful -- box lengths of $L\approx
25\,\sigma$).  Three qualitatively different outcomes were observed:
($i$) At sufficiently low temperature the bilayer adopted a ``gel''
phase; ($ii$) within a more elevated temperature range a fluid phase
can be reached, \emph{provided} $w_\romc\gtrsim 0.8\,\sigma$; ($iii$)
at sufficiently high temperature a bilayer under zero tension always
fell apart.  Fluid and gel phases are already clearly distinct under
visual inspection (long range order in the gel phase and none in the
fluid).  Across the gel-fluid boundary \cite{footgl} we observed a
sharp increase in in-plane diffusion constant $D$, an abrupt decrease
in orientational order $S = \frac{1}{2}\langle 3 (\VECa_i \cdot
\VECn)^2 -1 \rangle_i$ (where $\VECa_i$ is the vector along the axis
of the $i^{\text{th}}$ lipid and $\VECn$ is the average bilayer
normal) and the emergence of a nonzero flip-flop rate $r_\romf$ (the
probability per unit time that a single lipid changes its monolayer).
Typical values for these parameters along the $k_\romB T = 1.1
\epsilon$ isotherm are $D = 0.06 - 0.03 \sigma^2/\tau$, $S = 0.5 -
0.8$ and $r_\romf = 2 - 90 \times 10^{-5} \tau^{-1} $.  For $w_\romc =
1.6\sigma$ and $k_\romB T = 1.1\epsilon$ we also measured the bilayer
rupture tension $\Sigma_\romr \approx 4$mN/m and the compressibility
modulus at zero tension, $\mathcal{K} \approx 50$mN/m (mapping to real
length scales by assuming a bilayer thickness of 5nm). These
quantities are within their respective measured ranges for synthetic
and biological membranes, $\Sigma = 0.1-12$mN/m \cite{MoHo01} and
$\mathcal{K} = 50 - 1700$mN/m (depending on cholesterol content)
\cite{NeedNunn:90}. Second, lipids were simulated at constant volume,
but starting from a random ``gas'' configuration.  Under all
conditions which previously gave stable tensionless membranes, a
bilayer patch quickly self-assembled which at the right $L$ could zip
up with its open ends to span the box.  If the box was too big, the
patch either remained free, or (sometimes) closed upon itself to form
a vesicle.  Box spanning bilayers could also occur somewhat above the
evaporation boundary of Fig.~\ref{fig:pd} \cite{Brannigan_ensemble},
indicating that this line should be viewed as the location where the
rupture tension for a bilayer approaches zero.  We finally remark that
upon approaching this line from below, the bilayer becomes
increasingly disordered, and flip-flop rates and diffusion constants
increase strongly.


Looking at figure \ref{fig:pd} we see that the temperature range
over which a fluid bilayer is stable increases as $w_\romc$ is
increased and that for relatively narrow potentials $w_\romc <
0.7\sigma$ the fluid region disappears completely \cite{ljfail}.
It is noteworthy that these general features are not restricted to
the present functional form of Eqn.~(\ref{eqn:pot}).  Indeed, we have
also obtained a qualitatively similar phase diagram for a
Lennard-Jones like potential with variable width.  It should
therefore be emphasized that it is not the precise functional form
of our tail attractions that is important for the stabilization of
fluid bilayers but rather the length scale over which these
attractions are effective.

Next we illustrate that the fluid bilayers approach the correct
elastic continuum limit.  If one expands the bilayer in modes
$h(\VECr)=\sum_\VECq h_\VECq \, \rome^{\romi\VECq\cdot\VECr}$ with
$\VECq=\frac{2\pi}{L}(n_x,n_y)$, (linearized) Helfrich theory predicts
the mode spectrum \cite{Sei:97}
\begin{equation}
  \langle |h_\VECq^2|\rangle = \frac{k_\romB T}{L^2[\kappa q^4+\sigma q^2]} \ ,
  \label{eq:hq2}
\end{equation}
where $\kappa$ is the bending modulus and $\sigma$ the lateral
tension.  Below the crossover wave vector $q_\romc =
\sqrt{\sigma/\kappa}$ one has $\langle |h_\VECq^2|\rangle\sim
q^{-2}$ (tension regime), while $\langle |h_\VECq^2|\rangle\sim
q^{-4}$ holds above (bending regime).  Once $1/q$ approaches length
scales comparable to the bilayer thickness, continuum theory breaks
down and further effects (\eg\ protrusion modes \cite{LiGr93}) set in.
In order to see the characteristic $q^{-4}$ scaling of the bending
regime over a sufficiently wide range requires $q_\romc$ to be as
small as possible, hence we simulated at zero tension.  Furthermore,
to get away from microscopic lengths (such as the bilayer thickness)
we took systems four times as big as the ones we used for mapping the
phase diagram (4000 lipids, $L\simeq 50\,\sigma$).  Note that reaching
the continuum limit in MD simulations is not trivial, since the
relaxation time of bending modes scales as $q^{-4}$.

\begin{figure}
\includegraphics[scale=0.85]{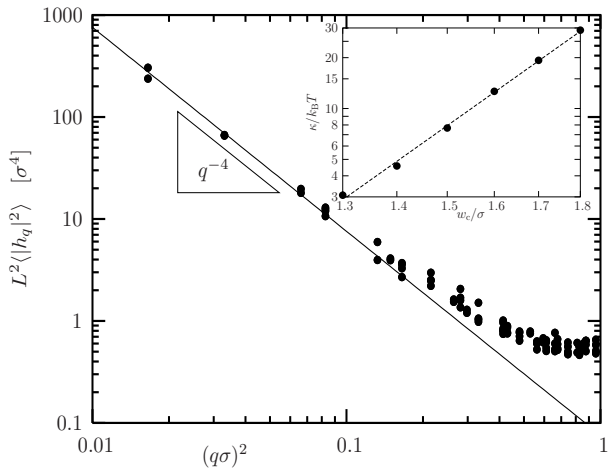}
\caption{Asymptotic $q^{-4}$ scaling of the power spectrum
  $\langle |h_\VECq^2|\rangle$ for the bilayer system with
  $w_\romc=1.6\sigma$ and $k_\romB T=1.1\epsilon$.  The inset shows the
  thus measured bending stiffness values at fixed temperature
  ($k_\romB T = 1.1 \epsilon$) as a function of potential range
  $w_\romc$.}\label{fig:kappa}
\end{figure}

After setting up the bilayer, we first waited until tension, box
length, and energy had equilibrated (which took typically $10^5
\tau$ for fluid systems).  Then on the order of 100 configurations
separated by $2000\tau$ were used to measure the mode spectrum. The
bilayer mid-plane was identified by tracking the tail-beads and
interpolating their vertical position onto a $16\times 16$ grid.
Possible stray lipids had to be excluded from this procedure.  An
FFT then yields the power spectrum $\langle |h_\VECq^2|\rangle$.
Fig.~\ref{fig:kappa} illustrates (for the system with
$w_\romc=1.6\sigma$ and $k_\romB T=1.1\epsilon$) the $q^{-4}$
scaling.  Notice that length scales exceeding $L\approx
20\,\sigma$ (\ie, about four times the bilayer thickness) are
required to reach the asymptotic regime, outside of which a fit to
Eqn.~(\ref{eq:hq2}) and the subsequent extraction of a bending
constant is meaningless.  The inset shows the corresponding values
of $\kappa$ for the six simulated stable bilayer systems at this
particular value of the temperature (see Fig.~\ref{fig:pd}).  We
emphasize that its value is easily tuneable over the
experimentally interesting range, at least from $\kappa \approx
3\,k_\romB T$ up to $\kappa \approx 30\,k_\romB T$.

\begin{figure}
\includegraphics[scale=0.2]{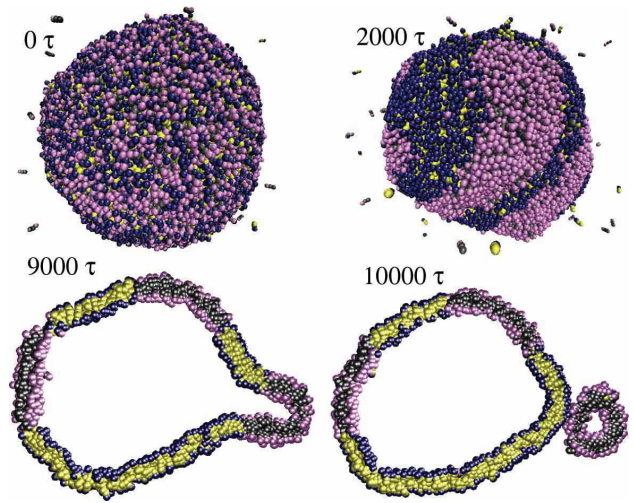}
\caption{Phase separation and budding sequence for a mixed vesicle
pre-assembled from $8000$ A-lipids and $8000$ B-lipids at $k_\romB
T=1.1 \epsilon$.  After an equilibration time of $2000\,\tau$, during
which $w_\romc^{\text{AA}}=w_\romc^{\text{BB}}=
w_\romc^{\text{AB}}= 1.5\sigma$, the cross term was reduced to
$w_\romc^{\text{AB}} = 1.3\sigma$.  Times indicated are measured
from this point on.}\label{fig:bud}
\end{figure}

So far we have used the parameter $w_\romc$ only to tune the bilayer
stiffness.  Let us now illustrate another application, namely, the
possibility of creating mixed lipid systems.  We consider a simple
example in which two lipid types A and B are present and where
$w_\romc^{\text{AA}} = w_\romc^{\text{BB}}$.  A line tension can now
be generated by choosing the cross interaction $w_\romc^{\text{AB}}$
smaller than the homogeneous ones.  Provided a sufficiently small
cross term has been chosen, A and B domains will form.  Figure
\ref{fig:bud} illustrates qualitative features of this process for a
critical quench (A:B = 1:1).  First we note that matching domains
always appear on inner and outer leaflets.  As is typical for a
critical system, domains are initially elongated in shape and then
coarsen to form circular patches or stripes.  In this case the domain
size and A-B line tension are sufficient to induce budding
\cite{domain_budding}.  This process has recently been studied in
detail via DPD simulations \cite{YamHy:03}.  For matching lipid
proportions in outer and inner leaflets, budding was induced by
cleavage of the domain boundary.  We also observed this ``flaking
mechanism'' for very high line tensions ($w_\romc^{\text{AB}} \ll
w_\romc^{\text{AA,BB}}$), while for less pronounced line tensions a
more continuous type budding occured (see Fig.~\ref{fig:bud}),
provided the vesicle was big enough.

Next we investigate the kinetics of domain formation on a larger
vesicle in the off critical regime.  Although the kinetics of
domain formation has been extensively studied for supported
membranes, only a small number of recent simulations
\cite{LaKu:04,KuGoLi:01,YamHy:03} and experiments \cite{blob_exp}
have tackled the problem for free vesicles.  The importance of
these studies is underscored by the fact that many features of a
vesicular system are absent in a flat supported membrane.  These
include bending fluctuations, volume constraints, and kinetic
effects of curvature-composition coupling \cite{Taniguchi}.

\begin{figure}
\includegraphics[scale=0.2]{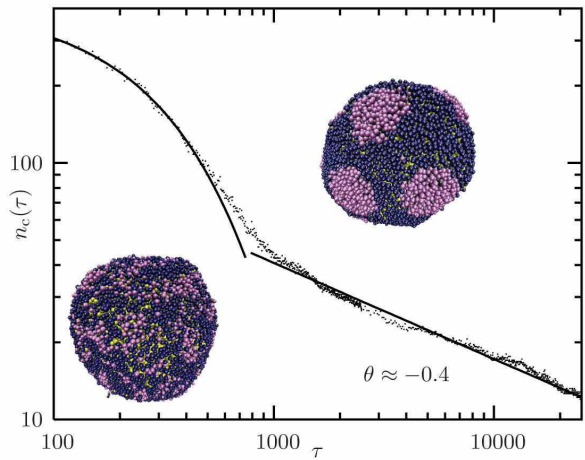}
\caption{Coarsening kinetics for a vesicle composed of $4800$
A-lipids and $11200$ B-lipids at $k_\romB T = 1.1\epsilon$ with
self interactions $w_\romc^{\text{AA}}=w_\romc^{\text{BB}} =
1.7\sigma$.  A cross term of $w_\romc^{\text{AB}} = 1.5\sigma$ was
imposed at $t = 0 \tau$.  The number $n(t)$ of clusters of size
three or greater was calculated as an average over five
independent runs.  After an initial exponential decrease, a power
law with exponent $\theta\approx -0.4$ is
observed.}\label{fig:kin}
\end{figure}

To study the kinetics of phase separation we performed simulations
using pre-equilibrated vesicles with an A:B ratio of 3:7 and measured
the number $n(t)$ of clusters of A lipids as a function of time. This
quantity displays two distinct kinetic regimes (see
Fig.~\ref{fig:kin}).  In the first regime $n(t)$ decays exponentially,
corresponding to a conversion of an initially exponential cluster size
distribution to an arrangement where meso-sized clusters begin to
dominate.  At later times such clusters display an $n(t) \sim
t^{\theta}$ power law decay with $\theta \approx -0.4$, which agrees
with the value expected for coarsening via patch-coalescence in the
non-hydrodynamical regime \cite{BinderPRL}.  This scenario is
confirmed visually, as well as by characteristic jumps in the (size
weighted) average cluster size. Interestingly, we do not observe
evidence for a Lifshitz-Slyozov type ripening, for which one would
have $\theta = -2/3$ \cite{Huse}.




Having demonstrated the key features of the present model, including
its tunability and its application to multi-component systems, we now
turn to performance aspects.  Although one of the clear advantages of
our approach is speed, it is difficult to make meaningful comparisons
across different computer architectures and implementations.  One
basic quantity that provides a rough implementation-independent
comparison is the particle number.  Simulating a vesicle with $16000$
coarse grained lipids, as in our example, would require roughly $25$
times as many solvent particles \cite{LaKu:04,YamHy:03}.  (Actually
this factor scales with the vesicle radius).  Obviously, one must also
account for the time-step which is often chosen somewhat larger in DPD
simulations.  Still, this leads us to an at least 5-fold speed-up for
the present method over DPD, and a crude comparison based on direct
CPU time usage agrees with this estimate \cite{cpuest}.

In summary, we have presented a method that is fundamentally
different from existing coarse grained lipid membrane simulations
that use explicit solvent.  Our method should be seen as
complementary to these techniques since it presents a compelling
speed advantage, especially for systems in three dimensions
(vesicles, bicontinuous phases).  It does not naturally include
volume constraints or hydrodynamics.  Indeed, the model is
inspired by the great success of simple ``bead-spring'' models
used to study polymer systems in the Rouse regime.  We have
presented Langevin Dynamics simulations here, but the simplicity
of the model and concept easily permit implementation of other
integrators or even  Monte Carlo.  Volume constraints should be
relatively simple and efficient to include via the concept of a
``phantom solvent'' \cite{LeSch04}.  The particle based nature of
the model and its tunable interactions allow for great flexibility
such that components with different pre-determined stiffness and
lipid shape present no difficulty, and topological changes occur
naturally.  Despite its astounding simplicity our approach allows
one to capture all these essential aspects of lipid bilayer
physics within a single unified framework.


We thank Oded Farago, Friederike Schmid, Hiroshi No\-gu\-chi, and
Bernward Mann for valuable discussions.




\begin{thebibliography}{99}
\bibitem{footgl} In the interest of brevity we have omitted a
detailed analysis of the gel-liquid transition and also of possible
subdivisions within the gel phase itself.  These issues will be
treated elsewhere.
\bibitem{Brannigan_ensemble}  This is most likely a finite-size effect 
(Brannigan, Tamboli, and Brown \cite{BrBr04} observe a similar
ensemble dependence).
\bibitem{ljfail} The LJ potential corresponds to $w_\romc\approx
  0.7\,\sigma$, and has indeed historically proven to be unsuccessful
  in creating fluid bilayers.
\end{thebibliography}
\end{document}